\documentclass[journal]{IEEEtran}

\usepackage{cite}
\usepackage{xurl}
\usepackage[hidelinks]{hyperref}
\usepackage{graphicx}
\usepackage[caption=false,font=footnotesize]{subfig}
\usepackage{cite}
\usepackage{amsmath}
\ifCLASSINFOpdf

\else
 
\fi

\usepackage{url}
\usepackage{bm}

\begin{document}
\bstctlcite{IEEEexample:BSTcontrol}

\title{Low Contrast Absorber Localization in Heat Pipe Microreactors Using Physics Informed Muon Tomography for Reactivity Monitoring}

\author{Reshma Ughade, and Stylianos Chatzidakis
        % <-this % stops a space
\thanks{R. Ughade and S. Chatzidakis are with the Department
of Nuclear Engineering, Purdue University, West Lafayette,
IN, 47907 USA (e-mail: ughaderashmi@gmail.com; schatzid@purdue.edu)}
% <-this % stops a space
% \thanks{J. Doe and J. Doe are with Anonymous University.}% <-this % stops a space
}

\maketitle

\begin{abstract}

Microreactors require nonintrusive monitoring methods that can provide information about internal control states without opening or modifying the sealed system. In heat pipe microreactors, rotating control drums regulate reactivity by changing the angular orientation of neutron absorber coatings around the reflector region. This study applies the previously developed \(\bm{\mu}\)TRec muon scattering tomography framework to absorber coating localization in the heat pipe microreactor geometry. Momentum informed \textbf{\(\bm{M}\)-value} imaging is used with the spectrometer information to evaluate whether the angular orientation of absorber coatings on twelve control drums can be inferred from reconstructed muon scattering images. Two absorber materials are considered. Hafnium is used as a high contrast reference case, while boron carbide, \textbf{\(\bm{B}_{\bm{4}}\bm{C}\)}, is used as the primary low contrast case because its density is close to that of the surrounding beryllium reflector. Both laser generated and cosmic ray muon sources are evaluated for absorber-in and absorber-out drum configurations. The reconstructed transverse images and one dimensional drum profiles show clear hafnium localization with both source types. For \textbf{\(\bm{B}_{\bm{4}}\bm{C}\)}, absorber localization is achieved with the controlled laser generated source when the number of events is increased, while the cosmic ray case remains limited by weak absorber contrast and artifacts near high scattering regions. The study defines the imaging conditions under which \(\bm{\mu}\)TRec based \textbf{\(\bm{M}\)-value} tomography can support control drum absorber localization for passive microreactor monitoring. These findings also support safeguards by design by informing the selection of absorber and reflector materials, dimensions, and coating thickness that improve nonintrusive detectability.

\end{abstract}


\IEEEpeerreviewmaketitle

\section{Introduction}

\IEEEPARstart{M}{icroreactors} are being developed as compact, factory-fabricated, and potentially transportable nuclear power systems for remote communities, industrial facilities, defense installations, and other applications requiring resilient and decentralized energy generation. Their anticipated deployment in remote or access-constrained locations, together with reduced on-site staffing and remote or semi-autonomous operation, increases the need for reliable monitoring methods that can provide independent information about the internal reactor state without opening or modifying the sealed system \cite{Stevens01122024,Ryan31072024,DOEMicroreactorProgram2025}. In heat-pipe microreactor concepts, passive heat transport is combined with rotating control drums positioned around the core and reflector. Rotation changes the orientation of neutron-absorbing material relative to the core, thereby controlling neutron absorption and core reactivity \cite{10.1115/ICONE28-67519}. Some microreactor designs also incorporate fail-safe mechanisms that drive the control drums toward a shutdown configuration following a protection signal or loss of electrical power \cite{Moon2023MARVEL}. Consequently, the actual angular position of each absorber coating provides important information about the reactor control state. A mechanical failure, actuator malfunction, or discrepancy between the commanded and actual drum positions could alter the expected reactivity response and create uncertainty in the available shutdown margin. A nonintrusive imaging method capable of independently verifying absorber positions could therefore complement conventional instrumentation by supporting reactor-state confirmation, fault diagnosis, and maintenance planning. Incorporating this capability during reactor development could also support safeguards by design by allowing absorber material, thickness, and geometry to be selected with nonintrusive inspectability and verification in mind \cite{IAEASafeguardsReactor2014}.

Cosmic ray muon imaging offers a promising nondestructive approach for characterizing dense, shielded, and otherwise inaccessible nuclear systems \cite{Borozdin_2003,SCHULTZ2004687,Miyadera:22, bae_gamma-ray_2024}. Because atmospheric muons are naturally available at the Earth’s surface and can penetrate thick layers of shielding, they provide a passive probe for imaging reactor components without requiring an external radiation source \cite{Borozdin_2003,Miyadera:22}. As muons pass through matter, they experience attenuation and multiple Coulomb scattering (MCS), both of which depend on the density, thickness, and radiation length of the materials along their path \cite{HIGHLAND1975497,LYNCH19916}. These properties have motivated the use of muon tomography for nuclear safeguards and monitoring applications, including spent nuclear fuel dry storage casks and other shielded nuclear objects \cite{chatzidakis_interaction_2016,bae_momentum_2024, chatzidakis_analysis_2016, liu_muon_2018}. Previous studies have shown that incorporating scattering information, and when available momentum measurements, can improve material discrimination and image reliability under limited muon statistics \cite{ughade_trec_2025, ughade_SciReport_2026, Ughade_assessment, ughade_physics-based_2023, ughade_--fly_2023, ughade_3d_2024, Ughade_efficient_2025, Ughade_resolution_2025, ughade_unsupervised_2026}. However, conventional reconstruction methods such as point of closest approach (PoCA) approximate the muon interaction as a single dominant scattering location \cite{SCHULTZ2004687, liu_muon_2018, Schultz_2007}. This assumption can reduce spatial accuracy in extended, curved, or heterogeneous geometries, where scattering occurs continuously along the muon path rather than at one isolated point.

In this work, the established muon trajectory reconstruction ($\mu$TRec) framework \cite{ughade_trec_2025, ughade_SciReport_2026, ughade_physics_2026} is used to investigate a new control state monitoring problem in heat pipe microreactors. Rather than detecting missing fuel or high density anomalies, the objective is to determine whether the angular orientation of control drum absorber coatings can be inferred from muon scattering images. This shifts the focus from bulk anomaly detection to a more localized and lower contrast image reconstruction task. The study first uses hafnium as a high contrast reference case to establish the expected imaging response when the absorber is clearly separated from the surrounding reflector. The analysis then focuses on B$_4$C, where the small density difference relative to beryllium makes absorber localization a low contrast imaging problem. Both laser based and cosmic ray muon sources are evaluated, and the absorber-in and absorber-out configurations are analyzed using reconstructed transverse images and one dimensional control drum profiles.

\section{Microreactor geometry}

\subsection{Heat pipe microreactor proxy}

\begin{figure}[h]
\centering

\subfloat[]{%
    \includegraphics[width= 1.1\columnwidth]{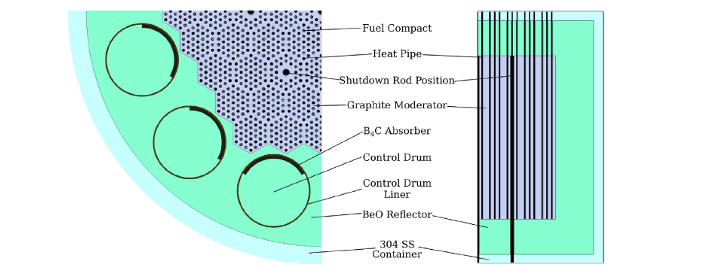}
    \label{fig:microreactor_geometry_a}
}

\vspace{2mm}

\subfloat[]{%
    \includegraphics[width=\columnwidth]{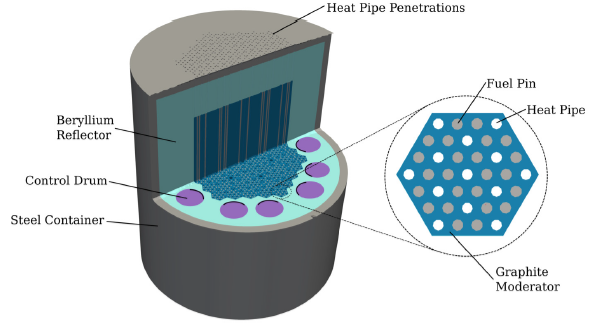}
    \label{fig:microreactor_geometry_b}
}

\caption{EMD microreactor geometry used in this study: (a) radial and axial core layout with labeled components \cite{PRICE2023112709} and (b) simulation layout with enlarged fuel flake \cite{PRICE2024134}.}
\label{fig:microreactor_geometry}
\end{figure}

The imaging object used in this study is a representative heat pipe microreactor geometry based on publicly available descriptions of the Westinghouse microreactor concept \cite{10.1115/ICONE28-67519}. The model is intended to capture the main structural and material features relevant to muon scattering tomography without relying on proprietary design information. The reactor consists of a graphite moderated core with UO$_2$ fuel, embedded heat pipes, shutdown rod locations, a surrounding reflector, and twelve rotating control drums. In the remainder of this paper, this geometry is referred to as the eVinci$^{TM}$ motivated design (EMD) microreactor. Fig.~\ref{fig:microreactor_geometry_a} shows the overall core layout, including the radial arrangement of the major components. Fig.~\ref{fig:microreactor_geometry_b} shows the fuel flake structure used in the simulation, where UO$_2$ fuel compacts and heat pipes are distributed within the graphite matrix. The principal geometric and material parameters used in the model are listed in Table~\ref{tab:emd_params}. The reference configuration does not include burnable absorbers, allowing the control drum absorber coatings to be isolated as the primary feature of interest in this study.

\begin{table}[!t]
\caption{Design parameters for the EMD microreactor. All pitches are reported as the flat-to-flat distances for the hexagonal array \cite{PRICE2023112709}.}
\label{tab:emd_params}
\centering
\footnotesize
\setlength{\tabcolsep}{3pt}
\renewcommand{\arraystretch}{1.08}
\begin{tabular}{p{0.55\columnwidth} c c}
\hline
\textbf{Parameter} & \textbf{Value} & \textbf{Unit} \\
\hline
Fuel pin diameter & 1.7 & cm \\
Active fuel height & 182 & cm \\
Drum liner outer radius & 18.7 & cm \\
Drum liner inner radius & 18.05 & cm \\
Drum liner material & lead & N/A \\
Drum absorber coating angle & 110 & $^\circ$ \\
Drum absorber coating thickness & 5 & cm \\
Drum height & 182 & cm \\
Heat pipe outer diameter & 1.6 & cm \\
Heat pipe wall thickness & 1.0 & mm \\
Heat pipe material & SS & N/A \\
Shutdown guide tube outer radius & 2.0 & cm \\
Shutdown guide tube thickness & 0.3 & cm \\
Shutdown guide tube material & SS & N/A \\
Pin cell pitch & 2.86 & cm \\
Flake pitch & 18 & cm \\
Reflector outer radius & 120 & cm \\
Container height & 280 & cm \\
Container outer radius & 130 & cm \\
Container thickness & 10 & cm \\
\hline
\end{tabular}
\end{table}

\section{Monte Carlo simulations}

\subsection{Muon source}

To separate the effect of source characteristics from the absorber reconstruction task, the simulations were performed using two incident muon configurations. The first configuration provides a controlled reference case. It uses a monoenergetic 5 GeV parallel muon beam directed toward the reactor model. Because the incident energy and direction are fixed, this case reduces source related variability and allows the reconstructed absorber response to be examined under simplified imaging conditions. This idealized source is motivated by recent high energy physics studies on laser generated directional muon beams for imaging and material inspection applications \cite{Terzani_2025, Calvin_2023, Ludwig_2025}

\begin{figure}[h]
\centering

\subfloat[Energy distribution]{%
    \includegraphics[width=0.7\columnwidth]{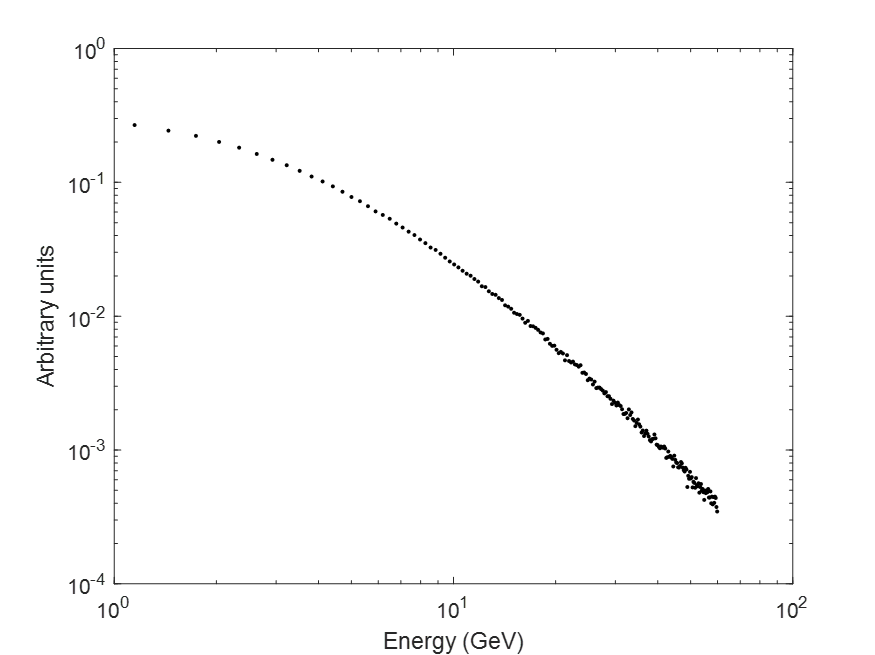}
    \label{fig:muon_energy}
}
\hfill
\subfloat[Angular distribution ]{%
    \includegraphics[width=0.7\columnwidth]{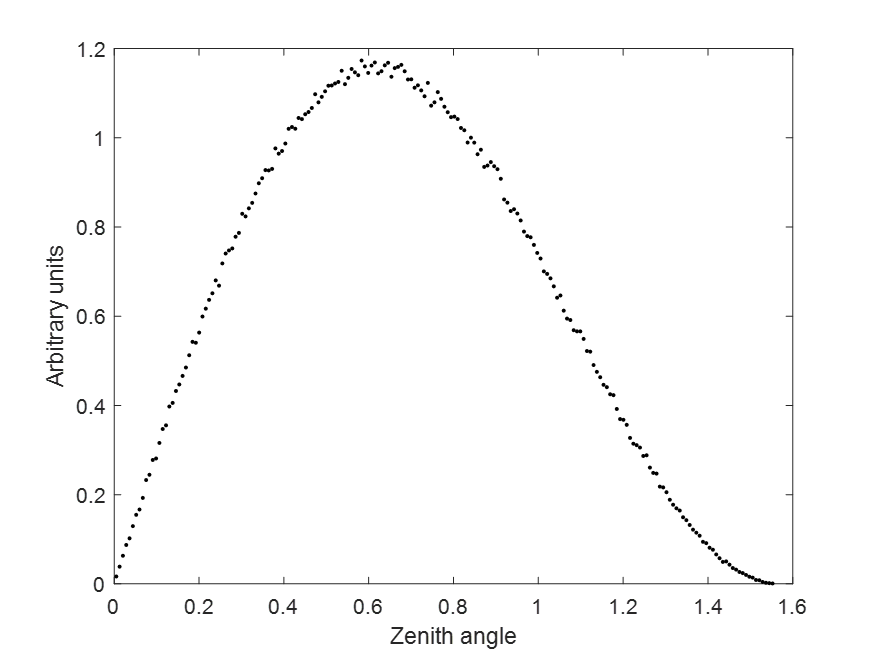}
    \label{fig:muon_zenith}
}

\caption{Atmospheric muon source used for the cosmic ray simulations in GEANT4: (a) sampled muon energy spectrum and (b) sampled zenith angle distribution \cite{chatzidakis_geant4-matlab_2015}.}
\label{fig:muon_source}
\end{figure}

The second configuration represents passive imaging with naturally available atmospheric muons. Cosmic ray events were generated in GEANT4 using the muon generator described in \cite{chatzidakis_geant4-matlab_2015}. In this case, muon energies range from 0 to 60 GeV, and the incident zenith angles range from 0 to 90$^\circ$. This configuration introduces the broad energy and angular spread expected in realistic cosmic ray measurements. The simulated cosmic ray energy and zenith angle distributions are shown in Fig.~\ref{fig:muon_source}.

\subsection{Detector and spectrometer configuration}

\begin{figure}[!t]
\centering

\subfloat[Geometry and detector setup.]{%
    \includegraphics[width=0.55\columnwidth]{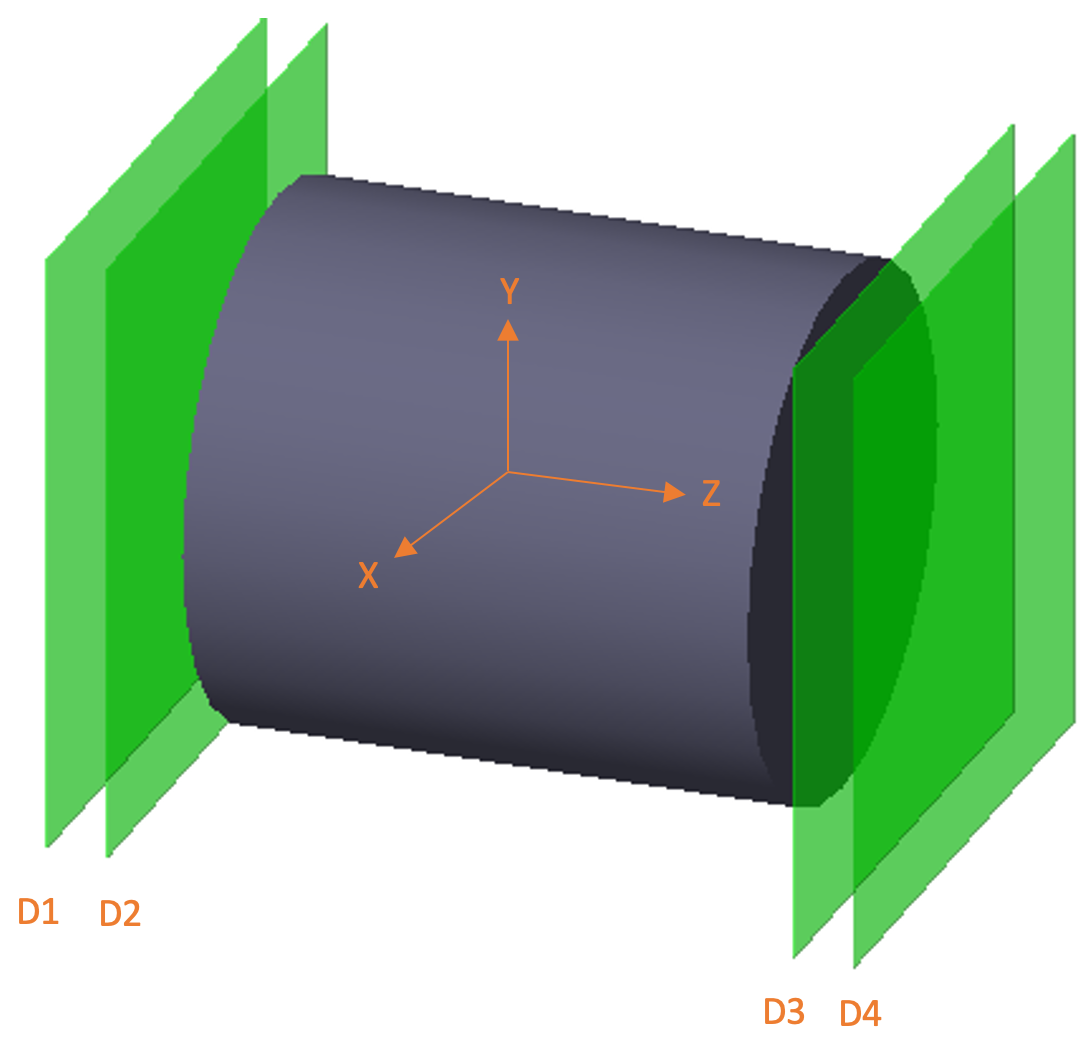}
    \label{fig:microreactor_setup}
}
\hfil
\subfloat[Side view]{%
    \includegraphics[width=0.40\columnwidth]{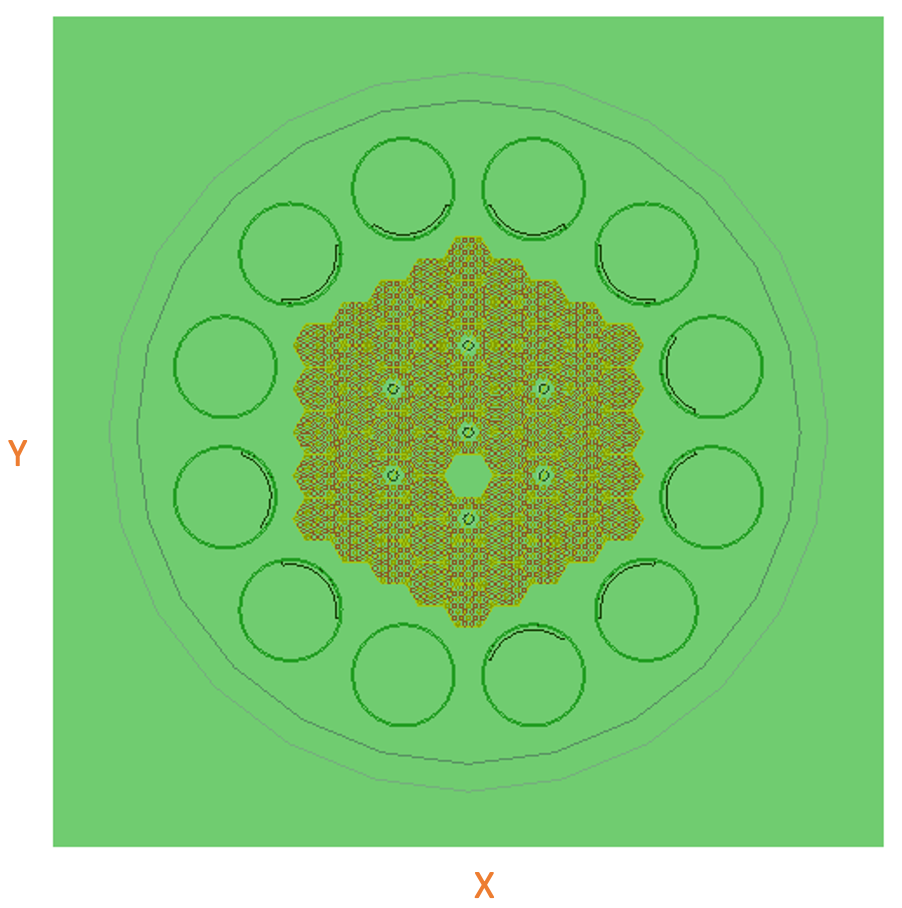}
    \label{fig:microreactor_side}
}

\caption{Measurement setup of the EMD microreactor modelled in GEANT4. The microreactor is shown in grey, and the four detector planes used for muon entry and exit tracking are shown in green \cite{ughade_SciReport_2026}.}
\label{fig:microreactor1}
\end{figure}

All particle transport simulations are performed in GEANT4 \cite{agostinelli_geant4simulation_2003}. The imaging setup consists of the EMD microreactor positioned between two tracking stations, as shown in Fig.~\ref{fig:microreactor_setup}. Each station contains two vertically oriented scintillation detector planes, giving four tracking planes in total. Each plane has dimensions of 300$\times$300$\times$1 cm$^3$. The distance between the first and fourth detector planes is 400 cm, and the spacing between the two planes within each tracking station is 30 cm.
The upstream detector pair records the incident muon track before the muon enters the microreactor geometry, while the downstream pair records the exiting track after the muon has passed through the core, reflector, and control drum regions. Only muons with valid position measurements in all four detector planes are retained for reconstruction. These four position measurements define the incoming and outgoing track segments used by the $\mu$TRec algorithm.
Momentum information is included using a two spectrometer configuration, with one momentum measurement on the entry side and one on the exit side. In the simulation, these measurements are represented using the muon information at detector D2 and detector D3, respectively. The entry side momentum is used in the $M$-value response, while the entry and exit momentum information together provide an estimate of energy loss through the object. This configuration is used for all reconstructions presented in this paper, allowing the absorber localization task to be evaluated with both trajectory and momentum information.

The same reconstruction and absorber orientation analysis can also be applied to a vertically oriented microreactor with horizontal detector planes placed above and below the reactor. In that arrangement, the detector geometry would make greater use of the near vertical cosmic ray muon flux, which is higher than the flux available for the horizontal reactor orientation considered here. Therefore, if detector placement below the microreactor is practical, the same method could be transferred directly to that configuration and would be expected to accumulate useful muon statistics more rapidly. Based on the increased vertical muon acceptance, the data collection rate is estimated to be approximately 235\% higher than in the horizontal configuration. The present study uses the horizontal microreactor orientation because it is more practical for deployment, installation, and external detector access.

\section{Method}

The reconstruction framework used in this study follows the previously developed $\mu$TRec muon scattering tomography method \cite{ughade_trec_2025, ughade_performance_2023, chatzidakis_generalized_2018, ughade2023muon, schulte_maximum_2008}. The full mathematical formulation of $\mu$TRec, including Gaussian multiple Coulomb scattering modeling, Bayesian trajectory estimation, momentum informed reconstruction, and voxelized image formation, is provided in \cite{ughade_SciReport_2026}. In the present work, the underlying trajectory estimation algorithm is not modified. Instead, the reconstructed $M$-value images are used for a new low contrast application, localizing the angular position of neutron absorber coatings in microreactor control drums.

For each accepted muon event $i$, the upstream and downstream detector pairs define the incident and exiting direction vectors, denoted by $\mathbf{u}^{\mathrm{in}}_i$ and $\mathbf{u}^{\mathrm{out}}_i$, respectively. The measured scattering angle is computed as
\begin{equation}
\theta_i =
\cos^{-1}
\left(
\frac{
\mathbf{u}^{\mathrm{in}}_i \cdot \mathbf{u}^{\mathrm{out}}_i
}{
\left|\mathbf{u}^{\mathrm{in}}_i\right|
\left|\mathbf{u}^{\mathrm{out}}_i\right|
}
\right).
\label{eq:scattering_angle}
\end{equation}

Using the measured entry and exit tracks, $\mu$TRec estimates the most probable muon trajectory through the microreactor volume. In each transverse projection, MCS is represented using a bivariate Gaussian model that jointly describes angular deflection ($\theta$) and lateral displacement ($y$). For the $y$--$z$ projection, the state vector is written as
\begin{equation}
\mathbf{Y}(z)=
\begin{bmatrix}
y(z)\\
\theta_y(z)
\end{bmatrix}.
\end{equation}
An analogous definition is used in the $x$--$z$ projection.

Energy loss is incorporated using the continuous slowing down approximation (CSDA), which is appropriate for minimum ionizing muons over the energy range relevant to this work. In contrast to straight line or single point scattering models, $\mu$TRec estimates a continuous curved trajectory that more faithfully represents the physical interactions experienced by muons in dense and nonuniform media. The reconstructed trajectory asymptotically approaches the measured incoming and outgoing tracks, which improves spatial localization and image fidelity.

\begin{figure}[htbp]
    \centering
    \includegraphics[width=0.7\linewidth]{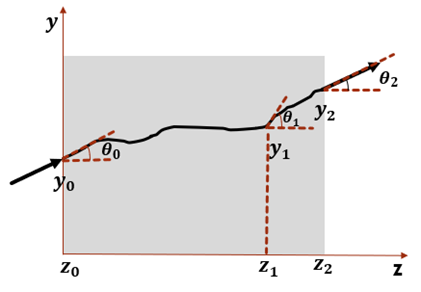}
    \caption{Schematic of multiple Coulomb scattering in the $y$--$z$ projection, showing the entry and exit track states and the scattering-induced displacement and angular deflection used in the $\mu$TRec formulation \cite{ughade_trec_2025}.}
    \label{muTRec_derivation}
\end{figure}

We consider the $y$--$z$ projection of the reconstruction volume, shown in Fig.~\ref{muTRec_derivation}, where $(y_0, z_0, \theta_0)$ denotes the muon state at the entry of the reconstruction volume and $(y_2, z_2, \theta_2)$ denotes the state at the exit. The $z$ direction is normal to the detector planes.

The $\mu$TRec estimate is written in compact form as
\begin{equation}
\mathbf{Y}_{\mu\mathrm{TRec}}(z)=\mathbf{A}^{-1}\mathbf{b}_Y ,
\label{eq:y_muTRec}
\end{equation}
where
\begin{equation}
\mathbf{A}=
\boldsymbol{\Sigma}_1^{-1}
+
\mathbf{R}_2^T
\boldsymbol{\Sigma}_2^{-1}
\mathbf{R}_2 ,
\label{eq:mutrec_A}
\end{equation}
and
\begin{equation}
\mathbf{b}_Y=
\boldsymbol{\Sigma}_1^{-1}
\mathbf{R}_0\mathbf{Y}_0
+
\mathbf{R}_2^T
\boldsymbol{\Sigma}_2^{-1}
\mathbf{Y}_2 .
\label{eq:mutrec_bY}
\end{equation}
Similarly, for the $x$--$z$ plane,
\begin{equation}
\mathbf{X}_{\mu\mathrm{TRec}}(z)=\mathbf{A}^{-1}\mathbf{b}_X ,
\label{eq:x_muTRec}
\end{equation}
with
\begin{equation}
\mathbf{b}_X=
\boldsymbol{\Sigma}_1^{-1}
\mathbf{R}_0\mathbf{X}_0
+
\mathbf{R}_2^T
\boldsymbol{\Sigma}_2^{-1}
\mathbf{X}_2 .
\label{eq:mutrec_bX}
\end{equation}
Here, $\Sigma_1$ and $\Sigma_2$ are covariance matrices for the upstream and downstream trajectory segments, respectively, and $R_0$ and $R_2$ are the corresponding transport matrices. Full definitions of these matrices, along with the derivation of Eqs.~\eqref{eq:y_muTRec} and \eqref{eq:x_muTRec}, are provided in Ref.~\cite{ughade_trec_2025}.

All reconstructions in this study use momentum information, each event is assigned a momentum dependent $M$-value \cite{bae_momentum_2024, bae_momentum-dependent_2022},
\begin{equation}
M_i =
\log_{10}
\left[
\left(
\frac{\theta_i}{0.958,p_i}
\right)^{2.4}
\right],
\label{eq:m_value_method}
\end{equation}
where $\theta_i$ is in radians and $p_i$ is the muon momentum in GeV/$c$. In the two spectrometer configuration used here, momentum information is available on both the entry and exit sides of the object, allowing the $\mu$TRec reconstruction to incorporate momentum dependent scattering and energy loss information.

Let $v$ denote a voxel and $\mathcal{I}(v)$ denote the set of reconstructed muon trajectories that pass through that voxel. The voxelized $M$-value response is computed as 
\begin{equation}
M(v) =
\frac{1}{|\mathcal{I}(v)|}
\sum_{i \in \mathcal{I}(v)}
M_i .
\label{eq:m_voxel}
\end{equation}

The resulting three dimensional $M$-value image is averaged along the axial direction of the microreactor to obtain a transverse reconstruction of the core, reflector, and control drum regions:
\begin{equation}
\bar{M}(x,y) =
\frac{1}{N_z}
\sum_{k=1}^{N_z}
M(x,y,z_k),
\label{eq:axial_average_method}
\end{equation}
where $N_z$ is the number of axial voxel layers. For visualization, the reconstructed $M$-value image is normalized and displayed using selected color limits. The display scale is used only for visualization and does not change the reconstructed voxel values.

\section{Absorber Material Selection and Imaging Contrast}

Control drums in heat pipe microreactors use neutron absorbing materials to regulate reactivity by changing the angular orientation of the absorber region relative to the core. When the absorber faces the core, neutron absorption increases and reactivity is reduced. When the absorber is rotated away from the core, fewer neutrons are absorbed and the reactor moves toward a more reactive configuration. Therefore, determining the angular position of the absorber coating provides a direct image based indicator of the control drum state. In this study, two absorber materials are considered: hafnium and boron carbide, B$_4$C. Both materials are modeled as 5 cm thick coatings to provide a consistent geometric basis for comparing their muon imaging response.

Hafnium is first investigated as a high contrast absorber case. Hafnium is attractive for reactor control applications because it has a large neutron absorption capability, good high temperature stability, and strong resistance to corrosion and radiation damage. From the perspective of muon tomography, hafnium is also favorable because its high density and high atomic number produce a strong multiple Coulomb scattering signature relative to the surrounding beryllium reflector. As a result, the hafnium coating is expected to appear as a clearly distinguishable high scattering region in the reconstructed image. This case is useful for validating whether the proposed $\mu$TRec based reconstruction can identify the absorber sector and recover the control drum orientation under favorable contrast conditions. However, hafnium also has practical limitations. It is relatively heavy and expensive, which can be undesirable for compact, transportable, and cost sensitive microreactor systems. 

B$_4$C is attractive for neutron control because boron, particularly $^{10}$B, has a very large neutron absorption cross section. Compared with hafnium, B$_4$C is lightweight and can provide strong neutronic control with a smaller mass penalty, making it suitable for compact microreactor designs. However, B$_4$C is significantly more difficult to localize using muon scattering tomography. Its density and effective scattering response are much closer to those of the beryllium reflector than hafnium. Therefore, the reconstructed contrast between the B$_4$C coating and the surrounding reflector is expected to be very weak.

\begin{figure}[h]
    \centering
    \subfloat[Absorber region]{%
        \includegraphics[width=0.7\columnwidth]{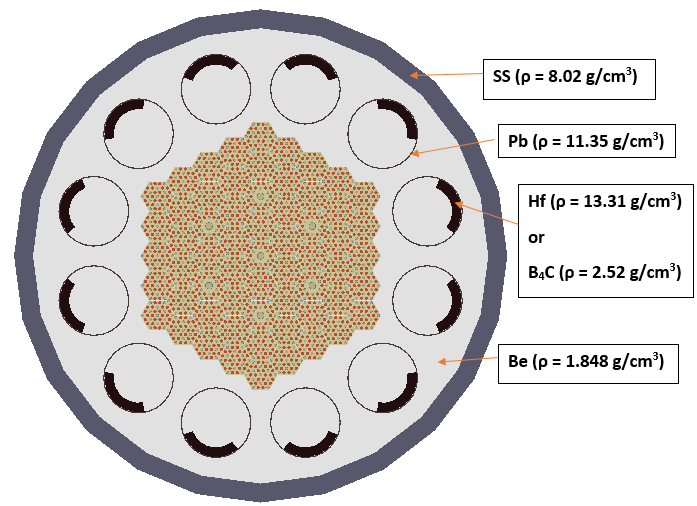}
        \label{fig:material_density1}
    }
    \hfill
    \subfloat[Core region]{%
        \includegraphics[width=0.6\columnwidth]{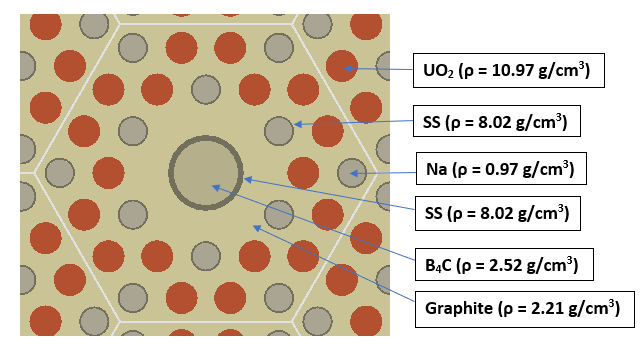}
        \label{fig:material_density2}
    }
    \caption{Reference density maps for interpreting absorber contrast and core material response in muon scattering tomography.}
    \label{fig:material_density}
\end{figure}

Fig.~\ref{fig:material_density1} illustrates the control drum absorber geometry with the corresponding material densities. Hafnium has a density of $\rho$ = 13.31 g/cm$^3$, which is much higher than the beryllium reflector density of $\rho$ = 1.848 g/cm$^3$. This large density difference produces a strong muon scattering signature, making the hafnium absorber coating relatively easy to localize. In contrast, B$_4$C has a density of $\rho$ = 2.52 g/cm$^3$, only 0.672 g/cm$^3$ higher than beryllium. This small density separation significantly reduces the scattering contrast between the absorber coating and the reflector background. Therefore, the B$_4$C case represents the more challenging imaging problem and motivates the use of additional image processing to enhance the weak absorber response. In addition, Fig.~\ref{fig:material_density2} provides a visual representation of the material density distribution in the microreactor geometry, which helps interpret the expected scattering response from different core and reflector components.

%%% Material density

\section{Results and discussion}

This section presents the reconstruction results for the two absorber materials considered in this study. First, the hafnium absorber case is evaluated using both the laser based muon source and the cosmic ray muon source. The same analysis is then repeated for the B$_4$C absorber case to assess the effect of reduced absorber to reflector contrast on control drum localization.

\subsection{Hafnium absorber}

\begin{figure}[h]
    \centering
    \subfloat[Absorber in]{%
        \includegraphics[width=0.48\columnwidth]{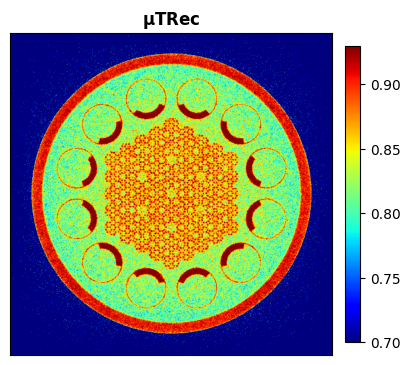}
        \label{fig:laser_hfa}
    }
    \hfill
    \subfloat[Absorber out]{%
        \includegraphics[width=0.48\columnwidth]{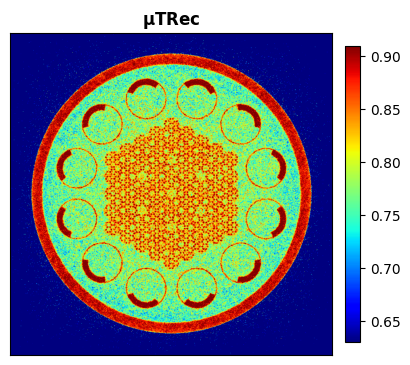}
        \label{fig:laser_hfb}
    }
    \caption{Hafnium absorber localization using $\mu$TRec with 3 million laser muon events and 4 mm voxels.}
    \label{fig:laser_hf}
\end{figure}

Fig.~\ref{fig:laser_hf} shows the reconstruction results for a 4 mm voxel size using 3 million laser muon events. The reconstructed values are first normalized and then displayed with adjusted color limits to emphasize the absorber-to-reflector contrast. In these images, the beryllium reflector appears in green, while the hafnium absorber coating is clearly visible in red. The hafnium coating shows strong contrast not only relative to the reflector, but also relative to the lead drum liner. Figs.~\ref{fig:laser_hfa} and \ref{fig:laser_hfb} show the absorber coating in the in and out drum orientations, corresponding to reduced and increased reactivity states, respectively. The clear separation between the hafnium absorber and the surrounding materials indicates that any mechanically mispositioned hafnium coating could be detected with high confidence using the $\mu$TRec algorithm with 3 million laser muon events. In addition, the internal core features, including the fuel pins, heat pipes, and shutdown rods, are clearly localized in the reconstructed images.

%%% Laser

\begin{figure}[h]
    \centering
    \subfloat[Line profiles]{%
        \includegraphics[width=0.68\columnwidth]{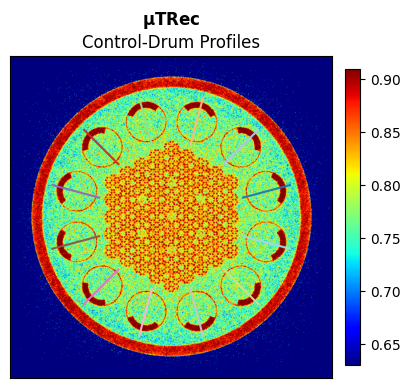}
        \label{fig:1Da}
    }
    \hfill
    \subfloat[Density profiles]{%
        \includegraphics[width=0.9\columnwidth]{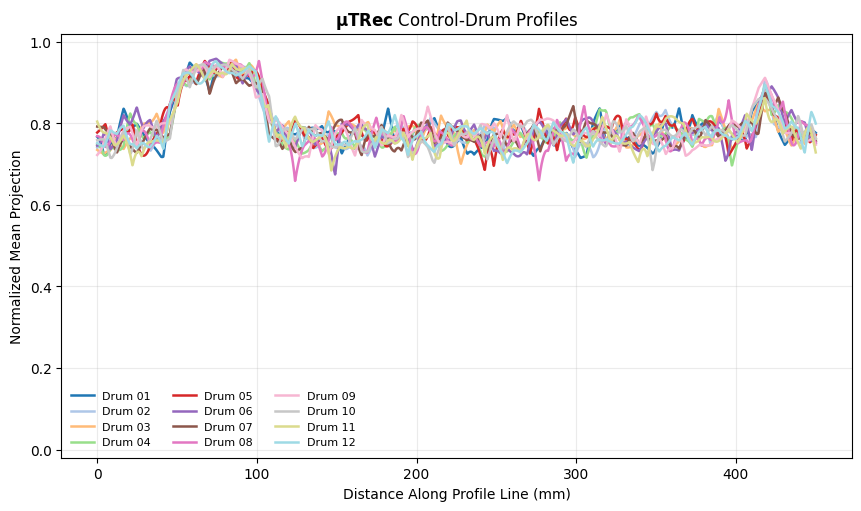}
        \label{fig:1Db}
    }
    \caption{One-dimensional density analysis of the hafnium absorber-out configuration along radial profiles through the twelve control drums.}
    \label{fig:1D}
\end{figure}

A one-dimensional representation of the reconstructed density along twelve radial profiles passing through the twelve control drums is shown in Fig.~\ref{fig:1D} for the absorber-out configuration. Fig.~\ref{fig:1Da} shows the locations of the selected line profiles, and Fig.~\ref{fig:1Db} presents the corresponding one-dimensional density profiles. Along each profile, the first peak corresponds to the combined high density contribution from the lead drum liner and the hafnium absorber coating, while the second peak corresponds primarily to the lead drum liner on the opposite side of the drum. The region without a distinct peak corresponds to the surrounding beryllium reflector. This clear peak structure demonstrates the strong contrast between the hafnium absorber, the lead drum liner, and the reflector background.

\begin{figure}[h]
    \centering
    \subfloat[Absorber in]{%
        \includegraphics[width=0.48\columnwidth]{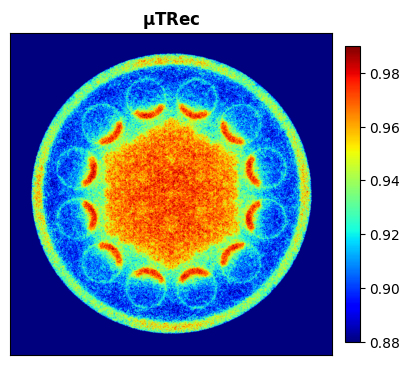}
        \label{fig:hf_cosmic1}
    }
    \hfill
    \subfloat[Absorber out]{%
        \includegraphics[width=0.48\columnwidth]{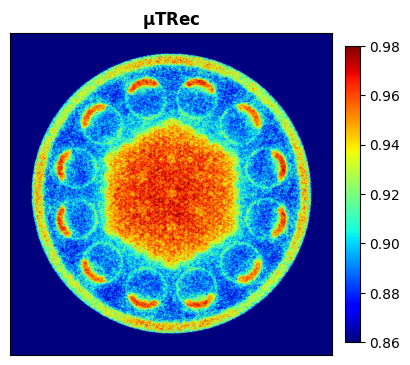}
        \label{fig:hf_cosmic2}
    }
    \caption{Hafnium absorber localization using $\mu$TRec with 3 million cosmic muon events and 4 mm voxels.}
    \label{fig:hf_cosmic}
\end{figure}

The hafnium absorber case is next evaluated using the cosmic ray muon source. Unlike the laser based source, cosmic ray muons have broad energy and angular distributions, which introduce additional statistical variation and trajectory uncertainty in the reconstructed image. Fig.~\ref{fig:hf_cosmic} shows the $\mu$TRec reconstruction obtained using 3 million accepted cosmic muon events and a 4 mm voxel size. The absorber-in configuration, shown in Fig.~\ref{fig:hf_cosmic1}, and the absorber-out configuration, shown in Fig.~\ref{fig:hf_cosmic2}, are both clearly identified, indicating that hafnium absorber localization remains reliable under realistic cosmic ray illumination. Compared with the laser muon case, the cosmic-ray reconstruction shows increased noise and reduced image sharpness. As a result, fine core features such as fuel pins, heat pipes, and shutdown rods are more difficult to resolve, although the overall core region remains clearly localized. In these reconstructions, the reflector appears in blue, the hafnium absorber in red, and the lead drum liner in light blue. The green regions near high density structures correspond to reconstruction artifacts and local fluctuations caused by stronger muon scattering in these regions. These artifacts are attributed to the combined effects of the broad cosmic muon spectrum, limited angular sampling, voxelization, and approximations in the physics-informed $\mu$TRec trajectory model. Despite these limitations, the hafnium coating position is still recovered with high confidence.

%%%%% Cosmic

\subsection{$B_4C$ absorber}

\begin{figure}[h]
    \centering
    \subfloat[Absorber in]{%
        \includegraphics[width=0.48\columnwidth]{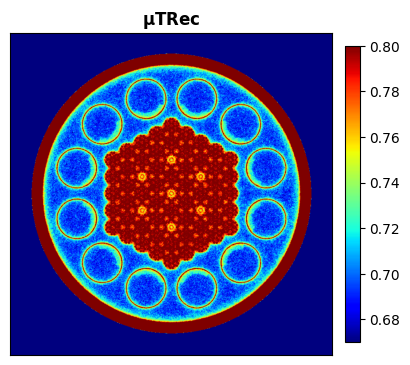}
        \label{fig:B4C_laser1}
    }
    \hfill
    \subfloat[Absorber out]{%
        \includegraphics[width=0.48\columnwidth]{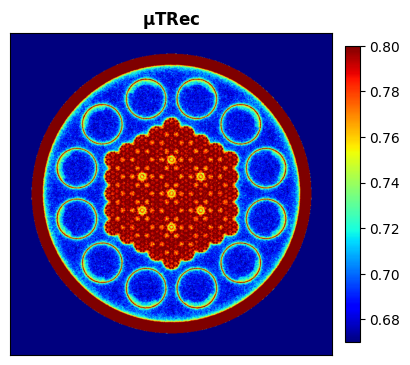}
        \label{fig:B4C_laser2}
    }
    \caption{B$_4$C absorber localization using $\mu$TRec with 20 million laser muon events and 4 mm voxels}
    \label{fig:B4C_laser}
\end{figure}

A similar analysis is performed for the B$_4$C absorber case. For this reconstruction, the display scale is adjusted to emphasize the absorber, drum, and reflector regions rather than the core features. Fig.~\ref{fig:B4C_laser} shows the absorber-in and absorber-out configurations reconstructed with a 4 mm voxel size for the 2.8 m tall microreactor geometry. Unlike the hafnium case, the density difference between the B$_4$C absorber and the beryllium reflector is only 0.672 g/cm$^3$. As a result, 3 million laser muon events are not sufficient to clearly separate the absorber coating from the surrounding reflector background. The number of muon events is therefore increased gradually, and the reconstruction obtained with 20 million laser muon events is presented. With this increased data set, the B$_4$C absorber positions can be successfully distinguished from the reflector and drum regions. Although the contrast remains much weaker than in the hafnium case, it is sufficient to localize the absorber positions with confidence.

%%% Laser

\begin{figure}[h]
    \centering
    \subfloat[Normalized scale]{%
        \includegraphics[width=0.47\columnwidth]{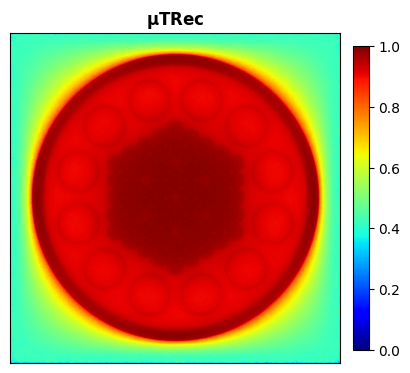}
        \label{fig:B4C_cosmic_abs_in1}
    }
    \hfill
    \subfloat[Absorber contrast]{%
        \includegraphics[width=0.49\columnwidth]{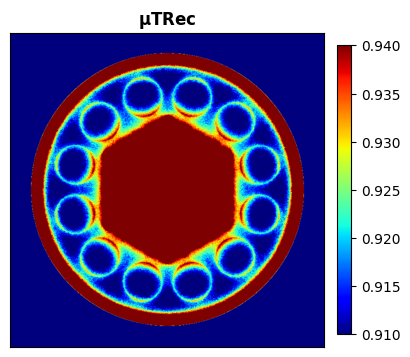}
        \label{fig:B4C_cosmic_abs_in2}
    }
    \hfill
    \subfloat[Core contrast]{%
        \includegraphics[width=0.47\columnwidth]{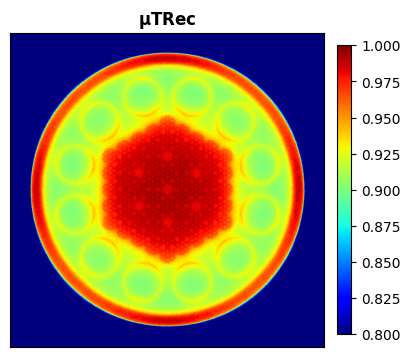}
        \label{fig:B4C_cosmic_abs_in3}
    }
    \hfill
    \subfloat[Core details]{%
        \includegraphics[width=0.49\columnwidth]{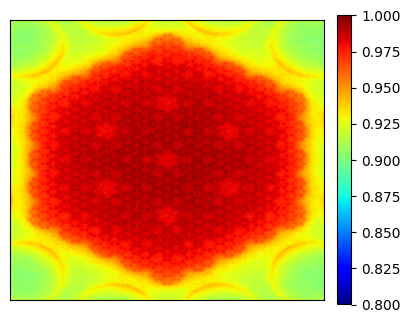}
        \label{fig:B4C_cosmic_abs_in4}
    }
    \caption{B$_4$C absorber-in reconstruction using $\mu$TRec with 20 million cosmic muon events and 4 mm voxels under different display scales.}
    \label{fig:B4C_cosmic_abs_in}
\end{figure}

Finally, the most practical and challenging case considered in this study is the B$_4$C absorber imaged using cosmic ray muons. Since the laser muon analysis required 20 million events to localize the low contrast B$_4$C coating, this case is expected to require greater statistics because of the additional noise introduced by the broad energy and angular distributions of cosmic muons. First, 20 million accepted cosmic muon events are analyzed. Fig.~\ref{fig:B4C_cosmic_abs_in} shows the reconstruction for the absorber-in configuration using 4 mm voxels. Fig.~\ref{fig:B4C_cosmic_abs_in1} presents the image with a normalized density scale, while Fig.~\ref{fig:B4C_cosmic_abs_in2} shows the same reconstruction with an adjusted scale chosen to emphasize the reflector-absorber-drum contrast. Under this display setting, the absorber position can be successfully identified, indicating that localization of the B$_4$C coating is achievable with 20 million cosmic muon events. Figs.~\ref{fig:B4C_cosmic_abs_in3} and \ref{fig:B4C_cosmic_abs_in4} show additional scale adjustments that place greater emphasis on the core geometry. Although these views are not optimized for absorber contrast, they still provide some indication of the absorber position. In addition, because the display is focused on the core region, the hexagonal fuel flakes become visible, with partial identification of fuel pins, heat pipes, and shutdown rods. These results show that even in the most difficult low contrast and realistic cosmic ray case, the B$_4$C absorber position can still be localized while retaining useful structural information about the microreactor core.

%% cosmic

\begin{figure}[h]
    \centering
    \subfloat[Absorber in]{%
        \includegraphics[width=0.48\columnwidth]{Results/B4C/cosmic/20M_4mm_in_final.png}
        \label{fig:B4C_cosmic1}
    }
    \hfill
    \subfloat[Absorber out]{%
        \includegraphics[width=0.48\columnwidth]{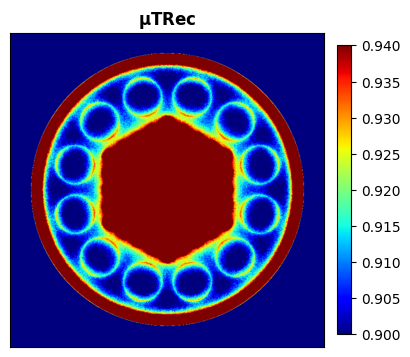}
        \label{fig:B4C_cosmic2}
    }
    \caption{B$_4$C absorber localization using $\mu$TRec with 20 million cosmic muon events and 4 mm voxels}
    \label{fig:B4C_cosmic}
\end{figure}

\begin{figure}[h]
    \centering
    \subfloat[Absorber in]{%
        \includegraphics[width=\columnwidth]{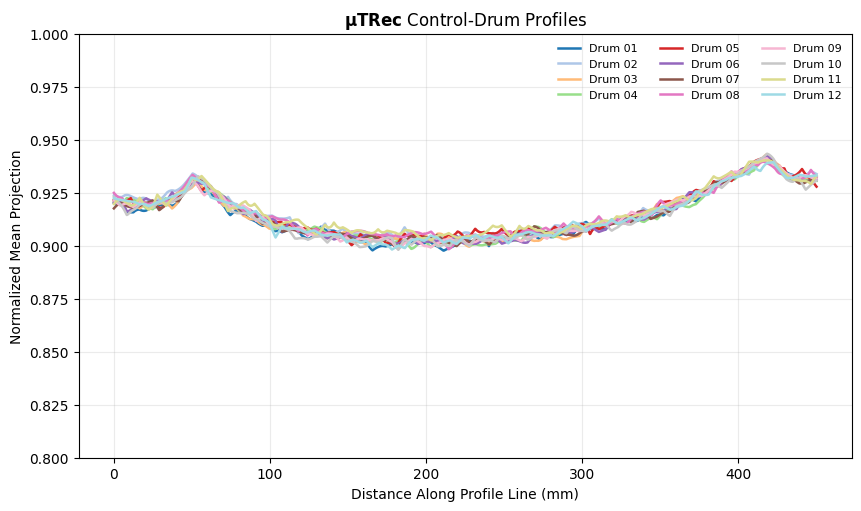}
        \label{fig:B4C_cosmic_1Da}
    }
    \hfill
    \subfloat[Absorber out]{%
        \includegraphics[width=\columnwidth]{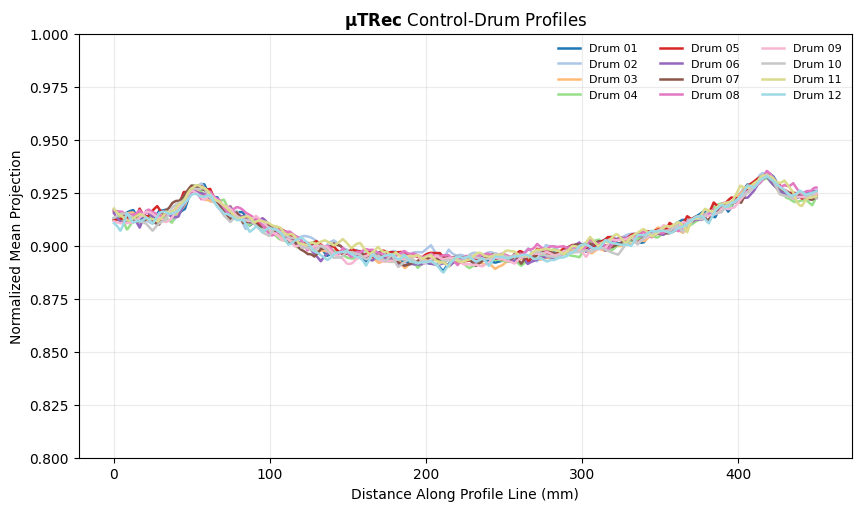}
        \label{fig:B4C_cosmic_1Db}
    }
    \caption{One-dimensional density analysis of the B$_4$C absorber-out configuration along radial profiles through the twelve control drums using 20 million cosmic muons.}
    \label{fig:B4C_cosmic_1D}
\end{figure}

However, the B$_4$C case with cosmicray muons remains the most challenging configuration. Fig.~\ref{fig:B4C_cosmic} shows the absorber-in and absorber-out reconstructions obtained using 20 million cosmic muon events and 4 mm voxels. Unlike the laser muon case, the cosmic ray reconstructions contain strong artifacts near the high density core and structural regions. As a result, the absorber-out configuration cannot be localized reliably. Although the absorber-in case appears to show red regions near the expected absorber locations in Fig.~\ref{fig:B4C_cosmic1}, this apparent localization cannot be considered conclusive because the same high response regions may be caused by reconstruction artifacts associated with strong scattering near the core. This limitation is further supported by the one-dimensional analysis across the twelve control drum profile lines, shown in Fig.~\ref{fig:B4C_cosmic_1D}. For a successful absorber orientation measurement, the absorber-out configuration in Fig.~\ref{fig:B4C_cosmic_1Db} is expected to produce a broader or stronger first peak, corresponding to the side of the profile that intersects the absorber coating. However, the profile appears similar to the absorber-in case. Similarly, the absorber-in configuration in Fig.~\ref{fig:B4C_cosmic_1Da} is expected to show a broader or stronger second peak, but the profiles do not show a clear shift between the two drum orientations. This indicates that, under the present reconstruction conditions, $\mu$TRec does not reliably distinguish the B$_4$C absorber position using 20 million cosmic muon events and 4 mm voxels.

These results highlight the need for further improvement before low contrast B$_4$C absorber localization can be performed reliably with cosmic ray muons. Future development should focus on reducing artifacts near high density regions, improving trajectory estimation for broad angle cosmic ray events, and incorporating more accurate material dependent scattering and energy loss information into the $\mu$TRec framework.

%%% Laser

\section{Conclusion}

This study investigated physics-informed muon tomography for nonintrusive localization of control drum absorber coatings in a heat-pipe microreactor. Two absorber materials were evaluated, hafnium as a high-contrast case and B$_4$C as a low-contrast case. For hafnium, $\mu$TRec successfully distinguished the absorber-in and absorber-out configurations using both laser-based and cosmic-ray muon sources. With 3 million events and 4 mm voxels, the absorber remained identifiable despite the higher noise observed in the cosmic-ray reconstructions. The B$_4$C case was more challenging because its density differs from that of the beryllium reflector by only 0.672 g/cm$^3$. With the laser-based source, increasing the data set from 3 million to 20 million events enabled localization of the absorber position. In contrast, the cosmic-ray source did not provide reliable localization, even when additional muon data were used. The reconstruction quality appeared to saturate, with high-density artifacts continuing to obscure the weak absorber signal and preventing clear separation of the two configurations in the one-dimensional profiles. This indicates that the limitation is not solely due to insufficient statistics, but also to the weak material contrast, cosmic-ray energy and angular variations, trajectory uncertainty, and reconstruction artifacts.
These results demonstrate that absorber detectability depends strongly on both the absorber material and its contrast relative to the surrounding reflector. The study therefore supports safeguards by design by showing how muon-imaging performance can be considered during reactor development. Absorber and reflector materials may be selected to provide sufficient scattering contrast, while absorber size and coating thickness may be designed to improve the nonintrusive verification of control component position. Future work should focus on artifact suppression, improved trajectory estimation, and reconstruction methods tailored to weak-contrast absorbers under cosmic-ray conditions.

\section*{Acknowledgment}
The present research is being conducted with the support of funding provided by the Purdue University School of Nuclear Engineering and the Purdue Research Foundation. 

\section*{Data availability}

Python and MATLAB scripts for $\mu$TRec and PoCA algorithms are publicly available at \url{https://github.com/rughade/muTRec-algorithm}. 

\ifCLASSOPTIONcaptionsoff
  \newpage
\fi

\bibliographystyle{IEEEtran}
\bibliography{references}

@IEEEtranBSTCTL{IEEEexample:BSTcontrol,
  CTLdash_repeated_names = "no"
}

@phdthesis{ughade_physics_2026,
  author = {Ughade, R. S.},
  title = {Physics Informed Muon Trajectory Reconstruction Algorithm for Enhanced Scattering Tomography},
  school = {Purdue University Graduate School},
  type = {{Ph.D.} dissertation},
  year = {2026},
  month = jul,
  note = {https://doi.org/10.25394/PGS.33098255.v1}
}

@techreport{IAEASafeguardsReactor2014,
  author      = {{International Atomic Energy Agency}},
  title       = {International Safeguards in the Design of Nuclear Reactors},
  institution = {IAEA Nuclear Energy Series},
  series      = {},
  number      = {NP-T-2.9},
  address     = {Vienna, Austria},
  year        = {2014},
  isbn        = {978-92-0-106514-8}
}

@techreport{Moon2023MARVEL,
  author      = {Brandon L. Moon},
  title       = {{TFR-2574 MARVEL Instrumentation and Control System (ICS)}},
  institution = {Idaho National Laboratory},
  number      = {INL/MIS-24-78446, Revision 0},
  address     = {Idaho Falls, ID, USA},
  year        = {2023},
  month       = {October}
}

@techreport{DOEMicroreactorProgram2025,
  author      = {{Idaho National Laboratory}},
  title       = {A Microreactor Program Plan for the Department of Energy},
  institution = {Idaho National Laboratory},
  number      = {INL/EXT-20-58919, Revision 4},
  address     = {Idaho Falls, ID, USA},
  year        = {2025},
  month       = {May}
}

@article{Ryan31072024,
author = {Emerald D. Ryan and Ishita Trivedi and Stefano Terlizzi and Ashley Shields and Mark Schanfein and Gustavo Reyes and Ryan Stewart},
title = {Challenges and Technology-Driven Opportunities for Safeguarding Microreactors},
journal = {Nuclear Science and Engineering},
volume = {0},
number = {0},
pages = {1--19},
year = {2024},
publisher = {Taylor \& Francis},
note = {https://doi.org/10.1080/00295639.2024.2372223},
}

@article{Stevens01122024,
author = {Kaeley Stevens and Joseph Oncken and Ronald Boring and Thomas Ulrich and Megan Culler and Haydn Bryan and Jeren Browning and Izabela Gutowska},
title = {Opportunities, Challenges, and Research Needs for Remote Microreactor Operations},
journal = {Nuclear Technology},
volume = {210},
number = {12},
pages = {2257--2273},
year = {2024},
publisher = {Taylor \& Francis},

}

@article{ughade_SciReport_2026,
    author = {Ughade, Reshma and Chatzidakis, Stylianos},
    title         = {Physics informed muon scattering tomography with momentum integration for nonintrusive monitoring of microreactor cores},
    journal       = {Scientific Reports},
    year          = {2026},
    note = {https://doi.org/10.1038/s41598-026-63149-z},
}

@article{ughade_trec_2025,
    author = {Ughade, Reshma and Chatzidakis, Stylianos},
	title = {{$\mu$TRec}: A Muon Trajectory Reconstruction Algorithm for Enhanced Scattering Tomography},
    journal = {Journal of Applied Physics},
	volume = {138},
    pages = {064909},
    number = {6},
    year = {2025},
	note = {https://doi.org/10.1063/5.0278370},
}

@article{ughade_performance_2023,
	title = {Performance Evaluation of Cosmic Ray Muon Trajectory Estimation Algorithms},
	volume = {13},
	note = {https://doi.org/10.1063/5.0174796},
	pages = {125301},
	number = {12},
	journal = {{AIP} Advances},
	author = {Ughade, Reshma and Bae, {JungHyun} and Chatzidakis, Stylianos},
	year = {2023},
}

@article{PRICE2023112709,
title = {Thermal Modeling of an {eVinci}\textsuperscript{TM}-Like Heat Pipe Microreactor Using {OpenFOAM}},
journal = {Nuclear Engineering and Design},
volume = {415},
pages = {112709},
year = {2023},
note = {https://doi.org/10.1016/j.nucengdes.2023.112709},
author = {Price, Dean and Roskoff, Nathan and Radaideh, Majdi I. and Kochunas, Brendan},
}

@article{PRICE2024134,
title = {Simplified Matching Pursuits Applied to {3D} Nuclear Reactor Temperature Distribution Construction},
journal = {Applied Mathematical Modelling},
volume = {131},
pages = {134-158},
year = {2024},
note = {https://doi.org/10.1016/j.apm.2024.04.011},
author = {Price, Dean and Radaideh, Majdi I. and Kochunas, Brendan},
}

@phdthesis{ughade2023muon,
  author       = {Reshma Sanjay Ughade},
  title        = {A Physics-Based Muon Trajectory Estimation Algorithm for Muon Tomographic Applications},
  school       = {Purdue University Graduate School},
  year         = {2023},
  type         = {Thesis},
  note = {https://doi.org/10.25394/PGS.23748933.v1},
}

@article{Ughade_assessment,
title = {Assessment of Performance for Algorithms Estimating Cosmic Ray Muon Trajectories},
author = {Reshma Ughade and Bae, {Jung Hyun} and Stylianos Chatzidakis},
year = "2023",
note = {\url{https://www.researchgate.net/publication/406190351_Assessment_of_Performance_for_Algorithms_Estimating_Cosmic_Ray_Muon_Trajectories}},
volume = "129",
pages = "369--372",
journal = {Transactions of the American Nuclear Society},
}

@inproceedings{ughade_physics-based_2023,
	title = {A Physics-Based Muon Trajectory Estimation Algorithm for Muon Tomographic Applications},
	volume = {2023},
	note = {https://ui.adsabs.harvard.edu/abs/2023APS..APRT16003U/abstract},
	booktitle = {{APS} {April} {Meeting} {Abstracts}},
	author = {Ughade, Reshma and Chatzidakis, Stylianos},
	year = {2023},
	pages = {T16--003},
}

@inproceedings{ughade_--fly_2023,
	title = {On-the-Fly {3D} Muon Path Estimation Algorithm for Muon Imaging Applications},
	note = {https://doi.org/10.1109/NSSMICRTSD49126.2023.10338661},
	booktitle = {2023 {IEEE} {Nuclear} {Science} {Symposium}, {Medical} {Imaging} {Conference} and {International} {Symposium} on {Room}-{Temperature} {Semiconductor} {Detectors} ({NSS} {MIC} {RTSD})},
	author = {Ughade, R. S. and Chatzidakis, S.},
	month = nov,
	year = {2023},
	pages = {1--1},
}

@incollection{bae_gamma-ray_2024,
	address = {Cham},
	title = {Gamma-Ray and Cosmic Ray Muon Modalities for Cargo Inspection},
	isbn = {978-3-031-63897-8},
	booktitle = {Emerging {Radiation} {Detection}: {Technology} and {Applications}},
	publisher = {Springer Nature Switzerland},
	author = {Bae, JungHyun and Chatzidakis, Stylianos and Ughade, Reshma},
	editor = {Iniewski, Kris and Gadey, Harish},
	year = {2024},
	note = {\url{https://doi.org/10.1007/978-3-031-63897-8_12}},
	pages = {207--238},
}

@inproceedings{ughade_3d_2024,
	title = {{3D} Analysis for Generalized Muon Trajectory Estimation Algorithm},
	volume = {131},
	note = {\url{https://www.researchgate.net/publication/406186275_3D_Analysis_for_Generalized_Muon_Trajectory_Estimation_Algorithm}}, 
	booktitle = {Transactions of the {American} {Nuclear} {Society}},
	author = {Ughade, Reshma and Bae, Jung Hyun and Cantonwine, Paul and Chatzidakis, Stylianos},
	year = {2024},
	pages = {856--859},
}

@inproceedings{ughade_unsupervised_2026,
	title = {Unsupervised Detection of High-Z Anomalies in Cargo Containers Using Muon Scattering Data},
	volume = {134},
	note = {\url{https://www.researchgate.net/publication/410865873_Unsupervised_Detection_of_High-Z_Anomalies_in_Cargo_Containers_Using_Muon_Scattering_Data}}, 
	booktitle = {Transactions of the {American} {Nuclear} {Society}},
	author = {Ughade, Reshma and Chatzidakis, Stylianos},
    month = {nov},
	year = {2026},
	pages = {184--187},
}

@article{chatzidakis_generalized_2018,
	title = {A Generalized Muon Trajectory Estimation Algorithm With Energy Loss for Application to Muon Tomography},
	volume = {123},
	note = {https://doi.org/10.1063/1.5024671},
	number = {12},
	journal = {Journal of Applied Physics},
	author = {Chatzidakis, Stylianos and Liu, Zhengzhi and Hayward, Jason P. and Scaglione, John M.},
	month = mar,
	year = {2018},
	pages = {124903},
}

@inproceedings{Ughade_efficient_2025,
  author={Ughade, R. S. and Chatzidakis, S.},
  booktitle={2025 IEEE Nuclear Science Symposium (NSS), Medical Imaging Conference (MIC) and Room Temperature Semiconductor Detector Conference (RTSD)}, 
  title={Efficient Low-Flux Muon Tomography Using the {$\mu$TRec} Algorithm}, 
  year={2025},
  pages={1-1},
  note = {\url{https://doi.org/10.1109/NSS/MIC/RTSD57106.2025.11287627}},
  }

@inproceedings{Ughade_resolution_2025,
  author = {Ughade, Reshma and Chatzidakis, Stylianos},
  title = {Integrating Detector Resolution into {$\mu$TRec} Algorithm for Spent-Fuel Cask Imaging},
  booktitle = {International High Level Radioactive Waste Management (IHLRWM)},
  year = {2025},
  volume = {129},
  pages = {61--64},
  note = {\url{https://www.researchgate.net/publication/400535806_Integrating_detector_resolution_into_TRec_algorithm_for_spent-fuel_cask_imaging}},
}

@article{Miyadera:22,
author = {Haruo Miyadera and Christopher L. Morris},
journal = {Appl. Opt.},
number = {6},
pages = {C154--C161},
publisher = {Optica Publishing Group},
title = {Muon Scattering Tomography: Review},
volume = {61},
month = feb,
year = {2022},
note = {https://doi.org/10.1364/AO.445806},
}

@article{liu_muon_2018,
	title = {Muon Tracing and Image Reconstruction Algorithms for Cosmic Ray Muon Computed Tomography},
	volume = {28},
	number = {1},
	journal = {IEEE Transactions on Image Processing},
	author = {Liu, Zhengzhi and Chatzidakis, Stylianos and Scaglione, John M. and Liao, Can and Yang, Haori and Hayward, Jason P.},
	year = {2018},
	pages = {426--435},
    note = {https://doi.org/10.1109/TIP.2018.2869667}
}

@article{chatzidakis_interaction_2016,
	title = {Interaction of Cosmic Ray Muons with Spent Nuclear Fuel Dry Casks and Determination of Lower Detection Limit},
	volume = {828},
	note = {https://doi.org/10.1016/j.nima.2016.03.084},
	journal = {Nuclear Instruments and Methods in Physics Research Section A: Accelerators, Spectrometers, Detectors and Associated Equipment},
	author = {Chatzidakis, S. and Choi, C. K. and Tsoukalas, L. H.},
	month = aug,
	year = {2016},
	pages = {37--45},
}

@article{chatzidakis_analysis_2016,
  author  = {Chatzidakis, S. and Choi, C. K. and Tsoukalas, L. H.},
  title   = {Analysis of Spent Nuclear Fuel Imaging Using Multiple Coulomb Scattering of Cosmic Muons},
  journal = {IEEE Transactions on Nuclear Science},
  volume  = {63},
  number  = {6},
  pages   = {2866--2874},
  month   = dec,
  year    = {2016},
}

@techreport{chatzidakis_geant4-matlab_2015,
	title = {A {Geant4}-{MATLAB} Muon Generator for {Monte Carlo} Simulations},
	author = {Chatzidakis, Stylianos},
    note = {https://doi.org/10.13140/RG.2.2.31871.41128},
	month = sep,
    institution = {Purdue University},
	year = {2015},
}

@article{bae_momentum-dependent_2022,
	title = {Momentum-{Dependent} {Cosmic} {Ray} {Muon} {Computed} {Tomography} {Using} a {Fieldable} {Muon} {Spectrometer}},
	volume = {15},
	note = {https://doi.org/10.3390/en15072666},
	number = {7},
	journal = {Energies},
	author = {Bae, Junghyun and Chatzidakis, Stylianos},
	month = jan,
	year = {2022},
	pages = {2666},
}

@article{bae_momentum_2024,
	title = {Momentum Informed Muon Scattering Tomography for Monitoring Spent Nuclear Fuels in Dry Storage Cask},
	volume = {14},
	note = {https://doi.org/10.1038/s41598-024-57105-y},
	number = {1},
	journal = {Scientific Reports},
	author = {Bae, JungHyun and Montgomery, Rose and Chatzidakis, Stylianos},
	month = mar,
	year = {2024},
	pages = {6717},
}

@article{schulte_maximum_2008,
	title = {A Maximum Likelihood Proton Path Formalism for Application in Proton Computed Tomography},
	volume = {35},
	note = {https://doi.org/10.1118/1.2986139},
	number = {11},
	journal = {Medical Physics},
	author = {Schulte, R. W. and Penfold, S. N. and Tafas, J. T. and Schubert, K. E.},
	year = {2008},
	pages = {4849--4856},
}

@article{agostinelli_geant4simulation_2003,
	title = {{Geant4}: A Simulation Toolkit},
	volume = {506},
	note = {https://doi.org/10.1016/S0168-9002(03)01368-8},
	number = {3},
	journal = {Nuclear Instruments and Methods in Physics Research Section A: Accelerators, Spectrometers, Detectors and Associated Equipment},
	author = {Agostinelli, S. and Allison, J. and Amako, K. and Apostolakis, J. and Araujo, H. and et al.},
	month = jul,
	year = {2003},
	pages = {250--303},
}

@article{HIGHLAND1975497,
title = {Some Practical Remarks on Multiple Scattering},
journal = {Nuclear Instruments and Methods},
volume = {129},
number = {2},
pages = {497-499},
year = {1975},
note = {https://doi.org/10.1016/0029-554X(75)90743-0},
author = {Highland, Virgil L.},
}

@article{SCHULTZ2004687,
title = {Image Reconstruction and Material {Z} Discrimination via Cosmic Ray Muon Radiography},
journal = {Nuclear Instruments and Methods in Physics Research Section A: Accelerators, Spectrometers, Detectors and Associated Equipment},
volume = {519},
number = {3},
pages = {687-694},
year = {2004},
note = {https://doi.org/10.1016/j.nima.2003.11.035},
author = {L.J. Schultz and K.N. Borozdin and J.J. Gomez and G.E. Hogan and J.A. McGill and C.L. Morris and W.C. Priedhorsky and A. Saunders and M.E. Teasdale},
}

@inproceedings{10.1115/ICONE28-67519,
  author = {Swartz, Matt M. and Byers, William A. and Lojek, John and Blunt, Rory},
  title = {Westinghouse {eVinci}\textsuperscript{TM} Heat Pipe Micro Reactor Technology Development},
  booktitle = {Proc. 28th International Conference on Nuclear Engineering (ICONE28)},
  year = {2021},
  pages = {V001T04A018},
  note = {https://doi.org/10.1115/ICONE28-67519},
}

@article{Borozdin_2003,
  title = {Radiographic Imaging with Cosmic-Ray Muons},
  author = {K.N. Borozdin and G. E. Hogan and C. Morris and W. C. Priedhorsky and A. Saunders and L. J. Schultz and M. E. Teasdale},
  journal = {Nature},
  volume = {422},
  pages = {277},
  year = {2003},
  publisher = {American Physical Society},
  note = {https://doi.org/10.1038/422277a},
}

@article{Terzani_2025,
  title = {Measurement of Directional Muon Beams Generated at the {Berkeley Lab Laser Accelerator}},
  author = {Terzani, Davide and Kisyov, Stanimir and Greenberg, Stephen and Le Pottier, Luc and Mironova, Maria and Picksley, Alex and Stackhouse, Joshua and Tsai, Hai-En and Li, Raymond and Rockafellow, Ela and Miao, Bo and Shrock, Jaron E. and Heim, Timon and Garcia-Sciveres, Maurice and Benedetti, Carlo and Valentine, John and Milchberg, Howard M. and Nakamura, Kei and Gonsalves, Anthony J. and van Tilborg, Jeroen and Schroeder, Carl B. and Esarey, Eric and Geddes, Cameron G. R.},
  journal = {Phys. Rev. Accel. Beams},
  volume = {28},
  number = {10},
  pages = {103401},
  year = {2025},
  month = oct,
  publisher = {American Physical Society},
  note = {https://doi.org/10.1103/kxjr-h7zs},
}

@article{Calvin_2023,
  author = {Calvin, Luke and Tomassini, Paolo and Doria, Domenico and Martello, Daniele and Deas, Robert M. and Sarri, Gianluca},
  title = {Laser-Driven Muon Production for Material Inspection and Imaging},
  journal = {Frontiers in Physics},
  volume = {11},
  pages = {1177486},
  year = {2023},
  note = {https://doi.org/10.3389/fphy.2023.1177486},
}

@article{Ludwig_2025,
  title = {Laser Based 100 {GeV} Electron Acceleration Scheme for Muon Production},
  author = {Ludwig, J. D. and Wilks, S. C. and Kemp, A. J. and Williams, G. J. and Lemos, N. and Rockafellow, E. and Miao, B. and Shrock, J. E. and Milchberg, H. M.},
  journal = {Scientific Reports},
  volume = {15},
  number = {1},
  pages = {25902},
  year = {2025},
  note = {https://doi.org/10.1038/s41598-025-95440-w},
}

@article{Schultz_2007,
  author={Schultz, Larry J. and Blanpied, Gary S. and Borozdin, Konstantin N. and Fraser, Andrew M. and Hengartner, Nicolas W. and Klimenko, Alexei V. and Morris, Christopher L. and Orum, Chris and Sossong, Michael J.},
  journal={IEEE Transactions on Image Processing}, 
  title={Statistical Reconstruction for Cosmic Ray Muon Tomography}, 
  year={2007},
  volume={16},
  number={8},
  pages={1985-1993},
  note = {https://doi.org/10.1109/TIP.2007.901239},
  }

@article{LYNCH19916,
title = {Approximations to Multiple Coulomb Scattering},
journal = {Nuclear Instruments and Methods in Physics Research Section B: Beam Interactions with Materials and Atoms},
volume = {58},
number = {1},
pages = {6-10},
year = {1991},
note = {https://doi.org/10.1016/0168-583X(91)95671-Y},
author = {Lynch, Gerald R. and Dahl, Orin I.},
}

\end{document}